\documentclass[a4paper,11pt]{article}
\pdfoutput=1 % if your are submitting a pdflatex (i.e. if you have
\usepackage{jheppub} % for details on the use of the package, please
\usepackage[T1]{fontenc} % if needed

\usepackage{amsmath,amssymb,tikz}
\usepackage{graphicx}
 \allowdisplaybreaks
\def\be{\begin{equation}}
\def\ee{\end{equation}}
\def\ba#1\ea{\begin{align}#1\end{align}}
\def\bg#1\eg{\begin{gather}#1\end{gather}}
\def\bm#1\em{\begin{multline}#1\end{multline}}
\def\bmd#1\emd{\begin{multlined}#1\end{multlined}}

\def\a{\alpha}
\def\b{\beta}

\def\d{\delta}

\def\e{\epsilon}

\def\l{\lambda}

\def\la{\label}

\def\({\left(}
\def\){\right)}
\def\[{\left[}
\def\]{\right]}

\def\Tr{{\rm Tr}}

\def \be {\begin{equation}}
\def \ee {\end{equation}}
\def \ba {\begin{array}}
\def \ea {\end{array}}
\def \bea{\begin{eqnarray}}
\def \eea{\end{eqnarray}}

\def \a {\alpha}
\def \b {\beta}

\def \d {\delta}

\def \e {\epsilon}

\def \l {\lambda}

\def \la {\leftarrow}
\def \ra {\rightarrow}

\def\bea{\begin{eqnarray}}
\def\eea{\end{eqnarray}}
\newcommand{\eq}[1]{(\ref{#1})}
\newcommand{\bit}{\begin{itemize}}  \newcommand{\eit}{\end{itemize}}
\newcommand{\ben}{\begin{enumerate}}  \newcommand{\een}{\end{enumerate}}

\def\la{\langle}
\def\ra{\rangle}

\def\cA{{\mathscr A}}

\def\Tr{{\rm Tr}}
\def\one{\mbox{1 \kern-.59em {\rm l}}}

\def\a{\alpha}        
\def\b{\beta}       
    
\def\d{\delta}    
\def\e{\epsilon}

\def\l{\lambda}

\def\cA{{\cal A}}

\def\ba{{\bf a}}

\def\d{\delta}

\def\uno{\mbox{1 \kern-.59em {\rm l}}}

\def\one{1\!\!1\,\,}

\def\bcomment#1{}
\def\IR{\relax{\rm I\kern-.18em R}}

\title{\boldmath Universality for Shape Dependence of Casimir Effects from Weyl Anomaly }

\author[a,1]{Rong-Xin Miao,\note{Corresponding author.}}
\author[a,b,2]{Chong-Sun Chu,}

\affiliation[a]{Physics Division, National Center for Theoretical Sciences,\\
National Tsing-Hua University, Hsinchu 30013, Taiwan}
\affiliation[b]{ Department of Physics, National Tsing-Hua
  University,
Hsinchu 30013, Taiwan}

\emailAdd{miaorongxin.physics@gmail.com}
\emailAdd{cschu@phys.nthu.edu.tw}

\abstract{ We reveal
elegant relations between the shape dependence of the
Casimir effects and Weyl anomaly in boundary conformal field theories 
(BCFT). We show that for any BCFT which has a description in terms of an
effective action, the near boundary divergent behavior of the 
renormalized stress tensor is completely determined by
the
central charges of the theory. These relations are verified by free BCFTs.
We
also
test them with holographic models of BCFT and find exact
agreement. We propose
that 
these relations between Casimir coefficients and central charges
hold for any BCFT.
With the holographic models, we reproduce  
not only the precise form of the near boundary divergent behavior of
the stress tensor, but also the surface counter term that is needed to
make the total energy finite. As they are
proportional to the central charges, the near boundary divergence of
the stress tensor must be physical and cannot be
dropped by further artificial renormalization.
Our results thus provide affirmative support on the
physical nature of the divergent energy density near the
boundary,
whose reality
has been a long-standing controversy in the
literature. }

\begin{document} 
\maketitle
\flushbottom

\section{Introduction}

The
 Casimir effect \cite{Casimir:1948dh}
 originates from the effect of boundary on the zero point energy-momentum
 of quantized fields in a system. As a fundamental
 %c5 energetic
 property of the
 quantum vacuum, it has important consequences on the  system of
 concern and has been applied to a wide range of 
 physical problems,
 %c4 from
 such as
 classic applications in the study of 
 %h1
 the
 Casimir force between conducting plates (and nano devices)
 \cite{Plunien:1986ca,Bordag:2001qi},
 %c2 QCD bag models \cite{Milton:1980ke}
 dynamical compactification of extra dimensions in string theory
 \cite{App1,App2},
 candidate of cosmological constant and dark  energy
 \cite{Milton:2004ya},
 as well as dynamical Casimir effect and its applications  \cite{dyn}.

The near boundary behavior of the stress tensor of a system is
crucial to the understanding of the Casimir effect. 
For a Quantum Field Theory (QFT) on a 
manifold $M$
%h1
of integer dimension $d$ and boundary $P$,
the renormalized  stress tensor is divergent
near the boundary \cite{Deutsch:1978sc}:
\begin{eqnarray}\label{stress0}
\la T_{ij} \ra  =
 x^{-d} T^{(d)}_{ij}...+x^{-1} T^{(1)}_{ij}, \quad  x \sim 0,
\end{eqnarray}
where $x$ is the proper distance from the boundary and
$T^{(n)}_{ij}$ with $n\ge 1$ depend only on the shape of the boundary
and the kind of QFT under consideration.
For CFT with conformal
invariant boundary condition (BCFT), one further require that
divergent parts of renormalized stress tensor are traceless in order
to get a well-defined finite Weyl anomaly without divergence. It is also
natural to impose the conservation condition of energy:
\begin{eqnarray}\label{stresstraceless}
  \lim_{x\to 0} \la T^i_{\ i} \ra =O(1),\quad \nabla_i \la T^i_{\ j} \ra
  =0.
\end{eqnarray}
%c1 In the following we will omit the expectation symbol $\la \; \ra$.
Substituting (\ref{stress0}) into the above equations,
\cite{Deutsch:1978sc} obtains
\begin{subequations}
  \label{solnT}
\begin{align}
& T^{(d)}_{ ij}=0, \quad \ T^{(d-1)}_{ij}=2 \a_1 \bar{k}_{ij}, 
\label{SolutionTij1} \\
& T^{(d-2)}_{ij}=\frac{-4\a_1}{d-1}n_{(i}h_{j)}^l\nabla_l
  k-\frac{4\a_1}{d-2}n_{(i}h_{j)}^ln^p R_{lp}\nonumber\\
&\ \ \  \ \ \ \  \ \ \ \  \ +\frac{2\a_1}{d-2}(n_i n_j-
  \frac{h_{ij}}{d-1})
  \Tr \bar{k}^2 + t_{ij}, \label{SolutionTij2} 
\end{align}
\end{subequations}
%cr1
\be
t_{ij} :=\lceil \b_1 C_{ikjl}n^kn^l+\b_2 \mathcal{R}_{ij}
  +\b_3 k k_{ij} +\b_4 k_i^lk_{lj} \rceil, \label{SolutionTij3} 
\ee
  where 
$n_i$, $h_{ij}$ and  $\bar{k}_{ij}$ are respectively 
the normal vector, induced metric and the traceless
part of extrinsic curvature of the boundary $P$.
The tensor $t_{ij}$ is tangential: $n^i t_{ij} =0$, 
$\lceil \ \rceil$ denotes the traceless part,
$C_{ijkl}$ is Weyl tensor of $M$ and $\mathcal{R}_{ij}$ is the intrinsic
Ricci tensor of $P$.  The coefficients $(\a, \b_i)$ fixes the shape
dependence of the leading and subleading Casimir effects of BCFT. The
main goal of this letter is to show that one can fix
completely these Casimir coefficients in terms of the bulk and boundary central
charges.

%%%%%%%%%%%%%%%%%%%%%%%%%%%%%%%%%%%
\section{Shape Dependence of Casimir effects from Weyl Anomaly}
%\section{Casimir effects from Weyl Anomaly}
%%%%%%%%%%%%%%%%%%%%%%%%%%%%%%%%%%%

Consider a BCFT
%c4 with a Lagrangian
with a well defined effective action.
The Weyl anomaly $\cA$,
%cr4 can be obtained
defined as the trace of
renormalized stress tensor,
can be obtained as the logarithmic
UV divergent term of the effective action,
%cr4
\be \label{I0}
I = \cdots + \cA \log (\frac{1}{\e})  + I_{\rm finite},
\ee
where $\cdots$ denotes terms which are UV divergent in powers of the
UV cutoff $1/\e$, and $ I_{\rm finite}$ is the renormalized, UV finite part of the
effective action. This part is dependent on the subtraction
scheme. But the dependence is irrelevant for the discussion below and
our results hold for any renormalization scheme.

Inspired by \cite{Lewkowycz:2014jia,Dong:2016wcf},
%cr9 we do not integrate over a small strip of geodesic distance $\epsilon$ from the
let us regulate  the effective action by excluding from its volume
integration a small  strip of  geodesic distance $\epsilon$ from the boundary.
Then there is no explicit boundary
divergences in this form of the effective action, however there are boundary
divergences implicit in the bulk effective action
which is integrated up to distance $\epsilon$.
%cr9 Then the
The variation of effective action is given by
\begin{eqnarray} \label{key1}
\delta I=  \frac{1}{2} \int_{x \ge \epsilon} \sqrt{g} \hat{T}^{ij}\delta g_{ij}
\end{eqnarray}
where $\hat{T}^{ij}=\frac{2 \delta I}{\sqrt{g}\delta g_{ij}}$ is the
non-renormalized bulk stress tensor.
%cr9 Now let us perform the standard step when people study
The renormalized bulk stress tensor is defined by the difference
of the  non-renormalized bulk stress tensor against a reference one
\cite{Deutsch:1978sc}:
\begin{eqnarray} \label{key3}
T^{ij}=\hat{T}^{ij}-\hat{T}_0^{ij},
\end{eqnarray}
where $\hat{T}_0^{ij}$ is the
non-renormalized stress tensor
defined for the same CFT without boundary. It is
\begin{eqnarray} \label{key2}
\delta I_{0}=  \frac{1}{2} \int_{x \ge \epsilon} \sqrt{g} \hat{T}_0^{ij}\delta g_{ij},
\end{eqnarray}
where $I_0$ is the effective action of  the  CFT 
with the boundary removed, hence the integration
over the region $x\ge \epsilon$.
Subtract (\ref{key2}) from (\ref{key1}) and focus on only the
logarithmically divergent terms, we obtain our key formula 
\begin{eqnarray} \label{key}
  (\delta \mathcal{A})_{\partial M} = \left( \frac{1}{2}
  \int_{x \ge \epsilon} \sqrt{g}T^{ij}\delta
g_{ij} \right)_{\log (1/\epsilon)},
\end{eqnarray}
where $(\delta \mathcal{A})_{\partial M}$ is the boundary terms in the variations
of Weyl anomaly and $T^{ij}$ is the renormalized bulk stress tensor.
In the above derivations, we have used the fact that $I$ and $I_0$ have
the same bulk Weyl anomaly so that 
\begin{eqnarray} \label{key3}
(\delta \mathcal{A})_{\partial M}= (\delta I-\delta I_0)_{\log (1/\epsilon)}.
\end{eqnarray}

%cr4
We observe that as the right hand side of \eq{key}
must give an exact variation, this  imposes strong constraints on
the possible form of the stress tensor near the boundary since this is
where one would pick up logarithmic divergent contribution on integration
near the boundary. It is this integrability of the variations 
which
%c3
helps us to fix the
Casimir effects in terms of the Weyl anomaly.
%cr4
To proceed, let us start with the metric written in the Gauss normal coordinates
\begin{eqnarray}\label{BCFTmetric0}
ds^2=dx^2+ \left(h_{ab}-2x k_{ab}+ x^2
q_{ab}+\cdots \right)dy^a dy^b,
\end{eqnarray}
where $x\in [0,+\infty)$.
  %cr1
  The coefficients $k_{ab}$, $q_{ab}$, $\cdots$ parametrize the
  derivative expansion
  %cr3
  (with respect to both $x$ and $y^a$)
  of the metric.
  %cr6
  Consider variation of the metric with
  $\delta g_{x i}=0$ and $\delta g_{ab}=\delta h_{ab}-2x \delta
  k_{ab}+ \cdots$.
  %cr6  where $\d h_{ab} = \nabla_a \xi_b+ \nabla_b \xi_a$,
  % $2 \d k_{ab} = \nabla_a \xi'_b + \nabla_b \xi'_a $, $\xi'_a
  % := \del_x \xi_a$. 
  %cr6 Consider
  Take first the 3d BCFT as an example.
  The Weyl anomaly of 3d BCFT is given by
\cite{Jensen:2015swa}
\begin{eqnarray}\label{3dBWA}
 \cA =\int_P\sqrt{h}(b_1 \mathcal{R}+b_2 \text{Tr}
  \bar{k}^2), 
\end{eqnarray}
where $b_1, b_2$ are boundary central charges which depends on the
boundary conditions. Taking the variation of (\ref{3dBWA}), we have
\begin{eqnarray}\label{3dBWAvariation1}
b_2 \int_{P}\sqrt{h}\Big[(\frac{ \text{Tr}
  \bar{k}^2}{2}h^{ab}-2 \bar{k}^{a}_c k^{c b})\delta
  h_{ab}+2\bar{k}^{ab}\delta k_{ab} \Big].
\end{eqnarray}
Now we turn to calculate the variation of Weyl anomaly from the last
term of (\ref{key}). Note that
%c4 we have
$C_{ijkl}=\lceil
\mathcal{R}_{ij} \rceil=0$  for $d=3$ . Note also that
$\bar{k}_{ij}(x)=g_i^{i'}g_j^{j'}\bar{k}_{i'j'}(0)=\bar{k}_{ij}(0)-2x
k_{(i}^l\bar{k}_{j)l}+O(x^2)$, where $g_i^{i'}$ is the bivector of
parallel transport between $x$ and $x=0$ \cite{Deutsch:1978sc}. Taking
these facts into account and substitute
\eq{stress0} and \eq{solnT}
%c1 \eq{SolutionTij1}, \eq{SolutionTij2}, \eq{SolutionTij3}
into the last term of (\ref{key}),
%c4 ;integration
integrate over $x$
and select the logarithmic divergent term, we obtain
\begin{eqnarray}\label{3dBWAvariation2}
&-&\a_1 \int_{P}\sqrt{h}[(\frac{ \text{Tr}
  \bar{k}^2}{2}h^{ab}-2 \bar{k}^{a}_c k^{c b})\delta
    h_{ab}+2\bar{k}^{ab}\delta k_{ab}] \nonumber\\  
&+& \int_{P}\sqrt{h}[(\frac{\b_3}{2}-\a_1) k \bar{k}^{ab}\delta
    h_{ab}+\frac{\b_4}{2} \lceil  k^a_ck^{cb}\rceil \delta h_{ab} ].
\end{eqnarray}
%cr4
Note that \eq{3dBWAvariation2} is made up
of
%cr5
a structure
of curvature components different from those
appearing in \eq{3dBWAvariation1}.
Integrability of \eq{3dBWAvariation2} gives $\b_3=2 \a_1$ and
$\b_4=0$.
%c5 Now we have fixed the stress tensor up to one parameter $\a_1$.
Comparing \eq{3dBWAvariation1} with \eq{3dBWAvariation2}
%cr4
gives $\a_1=-b_2$. All together, we obtain the relations between the
Casimir coefficients of the stress tensor and the boundary central charges:
\begin{eqnarray}\label{3dparameters}
\a_1=-b_2, \quad \b_3=-2 b_2, \quad \b_4=0.\;\;\;\;
\end{eqnarray}

Similarly for 4d BCFT, we can obtain the shape dependence of Casimir
effects from the Weyl anomaly \cite{Fursaev:2015wpa, Herzog:2015ioa}
\begin{eqnarray}\label{4dBWA}
  \cA &=&\int_M \sqrt{g}(\frac{c}{16\pi^2}C^{ijkl}C_{ijkl}
  -\frac{a}{16\pi^2} E_4)\nonumber\\
&&+\int_{P}\sqrt{h}( b_3 \text{Tr}
  \bar{k}^3 + b_4 C^{ac}_{\ \ \ b c}
  \bar{k}_{\ a}^b),\;\;
\end{eqnarray}
where $a, c$ are bulk central charges and $b_3, b_4$
are boundary central charges. $E_4$ is the Euler density including the
boundary term. To derive $t_{ij}$, 
%c1 (\ref{SolutionTij3}),
%newv1 we can turn off
we set
$\delta h_{ij}=0$ for simplicity, since it only
%cr3 matters for $O(k^3)$
affects the third order derivative  
terms in the stress tensor. Taking variation of (\ref{4dBWA}) and
comparing the boundary term with the last term of (\ref{key}),
we obtain
\be\label{4dparameters1}
\begin{array}{lll}
  \a_1=\frac{b_4}{2}, & \b_1=\frac{c}{2\pi^2}+b_4, &\b_2=0, \\
&   \b_3=2 b_3+\frac{13}{6}b_4, &  \b_4=-3 b_3-2 b_4. 
\end{array}
\ee
%newv2
%It is remarkable that 
%Note that the stress tensor depends on both the boundary
%and bulk central charges.
%cr2
It is remarkable that the boundary behavior of the stress tensor is completely 
determined by the boundary and bulk central charges
However, it is independent of the central
%c5 charges
charge related to Euler density
%c4 ,
due to the fact that topological
invariants do not change under local variations. We propose that the
relations \eq{3dparameters}
%c4 ,
and \eq{4dparameters1} between Casimir
coefficients and central charges
%c4 to
hold for general BCFT.

%%%%%%%%%%%%%%%%%%%%%%%%%%%%%%%%%%%
%c6
\section{Free and Holographic BCFT}
%%%%%%%%%%%%%%%%%%%%%%%%%%%%%%%%%%%

Let us verify our general statements with free BCFT. The
renormalized stress tensor of 4d free BCFT has been calculated in
\cite{Deutsch:1978sc,Kennedy:1979ar,Kennedy:1981yi}. The bulk and
boundary central charges for 4d free BCFTs
%c3 are
were obtained in
\cite{Fursaev:2015wpa}. We summary these results in Table \ref{table1}
and Table \ref{table2}. 
% newv6
Note that the
%cr2 date
results for
Maxwell field apply to both absolute and relative B.C. We find
these data obey exactly the relations
\eq{4dparameters1}.
%c1 , which is a strong support for our general statements.
%c4 change b1 to \b1
$\b_1$ for Maxwell field is absence in the
literature. Here from \eq{4dparameters1}, we
predict that $\b_1=0$ for all 4d free BCFT due to the fact that
$c=-2\pi^2 b_4$ for 4d free  BCFT. As we will show below, this relation
is violated by strongly-coupled CFT dual to gravity. As a result,
$\b_1$ is non-zero in general. Comparing with \cite{Kennedy:1981yi}, we
note that there is a minus sign typo of $\b_4$ for Maxwell field in
\cite{Deutsch:1978sc}.
% newv6

\begin{table}[ht]
\caption{Casimir coefficients for 4d free BCFT}
\begin{center}
    \begin{tabular}{| c | c | c | c |  c |  c|}
    \hline
     & $\a_1$ & $\b_1$ & $\b_2$  & $\b_3$ & $\b_4$\\ \hline
    Scalar, Dirichlet B.C   & $-\frac{1}{480 \pi ^2}$ & 0 & 0 &
    $-\frac{19}{10080 \pi ^2}$ & $-\frac{1}{420 \pi ^2}$ \\ \hline
    Scalar, Robin B.C     & $-\frac{1}{480 \pi ^2}$ & 0 & 0 &
    $ -\frac{1}{288\pi ^2}$ & 0 \\ \hline
    Maxwell field& $-\frac{1}{40 \pi ^2}$ &
    %c5 0
    (0)  & 0 &
    $ -\frac{43}{840\pi ^2}$  & $ \frac{1}{70\pi ^2}$  \\
    \hline
    \end{tabular}
\end{center}
\label{table1}
\end{table}

\begin{table}[ht]
\caption{Central charges for 4d free  BCFT}
\begin{center}
    \begin{tabular}{| c | c | c | c |  c | }
    \hline
     & $a$ & $c$ & $b_3$  & $b_4$ \\ \hline
  Scalar, Dirichlet B.C   & $\frac{1}{360}$ & $\frac{1}{120}$ & $\frac{1}{280 \pi ^2}$ & $-\frac{1}{240 \pi ^2}$  \\ \hline
  Scalar, Robin B.C     & $\frac{1}{360}$ & $\frac{1}{120}$ & $\frac{1}{360 \pi ^2}$ & $-\frac{1}{240 \pi ^2}$  \\ \hline
  Maxwell field & $\frac{31}{180}$ & $\frac{1}{10}$ & $\frac{1}{35 \pi ^2}$& $-\frac{1}{20 \pi ^2}$  \\
    \hline
    \end{tabular}
\end{center}
\label{table2}
\end{table}

%%%%%%%%%%%%%%%%%%%%%%%%%%%%%%%%%%%
%\section{Holographic BCFT}
%%%%%%%%%%%%%%%%%%%%%%%%%%%%%%%%%%%

Now let us investigate the shape dependence of Casimir effects in
holographic models of BCFT. Consider a BCFT defined on a manifold $M$
with a boundary $P$.
%h1 Recently,
Takayanagi \cite{Takayanagi:2011zk}
proposed to extend the $d$ dimensional manifold $M$ to a $d+1$
dimensional asymptotically AdS space $N$ so that $\partial N= M\cup
Q$, where $Q$ is a $d$ dimensional manifold which satisfies $\partial
Q=\partial M=P$. The gravitational action for holographic BCFT 
is \cite{Takayanagi:2011zk} ($16\pi G_N =1$)
\begin{eqnarray}\label{action1}
I&=&\int_N \sqrt{G} (R-2 \Lambda) +2\int_Q \sqrt{\gamma} (K-T) 
% c3 + 2\int_M\sqrt{g} K \nonumber\\
% c3 && + 2 \int_P \sqrt{\s} \th,
\end{eqnarray}
%c4
plus terms on $M$ and $P$.
%c4 a Gibbons-Hawking term on $M$ and a junction term on $P$.
Here $T$ is a constant which can be regarded as the holographic dual
of boundary conditions of BCFT
\cite{Miao:2017gyt,Chu:2017aab}.
A central issue in the construction of the AdS/BCFT 
is the determination of the location of $Q$ in the
bulk. \cite{Takayanagi:2011zk} propose to use the Neumann boundary
condition
\begin{eqnarray}\label{NBC}
K_{\alpha\beta}-(K-T)\gamma_{\alpha\beta}=0
\end{eqnarray}
to fix the position of $Q$.
In \cite{Miao:2017gyt,Chu:2017aab} we found there is generally no
solution to \eq{NBC} for bulk metric that
%c4 arised
arose from the FG expansion of a general non-symmetric boundary.
The reason is because $Q$ is of co-dimension one and
we only need one condition to determine it's position,
%cr1
while there are too many extra conditions in (\ref{NBC}).
To resolve this, we suggested in \cite{Miao:2017gyt,Chu:2017aab}
to use the trace of \eq{NBC}, $(1-d)K+dT=0$, to determine the position
of $Q$.
%cr1
Nonetheless, it is also possible that one may need to
relax the assumption that the bulk metric admits a valid FG expansion, as
has been attempted in \cite{Nozaki:2012qd} for some
non-symmetric boundary in BCFT${}_3$. 
In contrast to a FG-expanded metric whose form near the boundary $M$ is completely
fixed, a non-FG expanded metric has more degree of freedom. It
was suggested in  \cite{Nozaki:2012qd} that the embedding equation
\eq{NBC} may admit a solution if the bulk metric is also allowed to
adjust itself. 
%cr1
However
in general this is a highly non-trivial problem and
there is no systematic method available to construct gravity solutions for BCFT
in general dimensions $d$ and
with an arbitrary non-symmetric boundary ($\bar{k}_{ab}\ne 0$)
%cr1 which
that  is not FG expanded.
%cr1
Remarkably this problem can solved and we will now present the solution.

%h1
\begin{figure}[t]
\centering
\includegraphics[width=5cm]{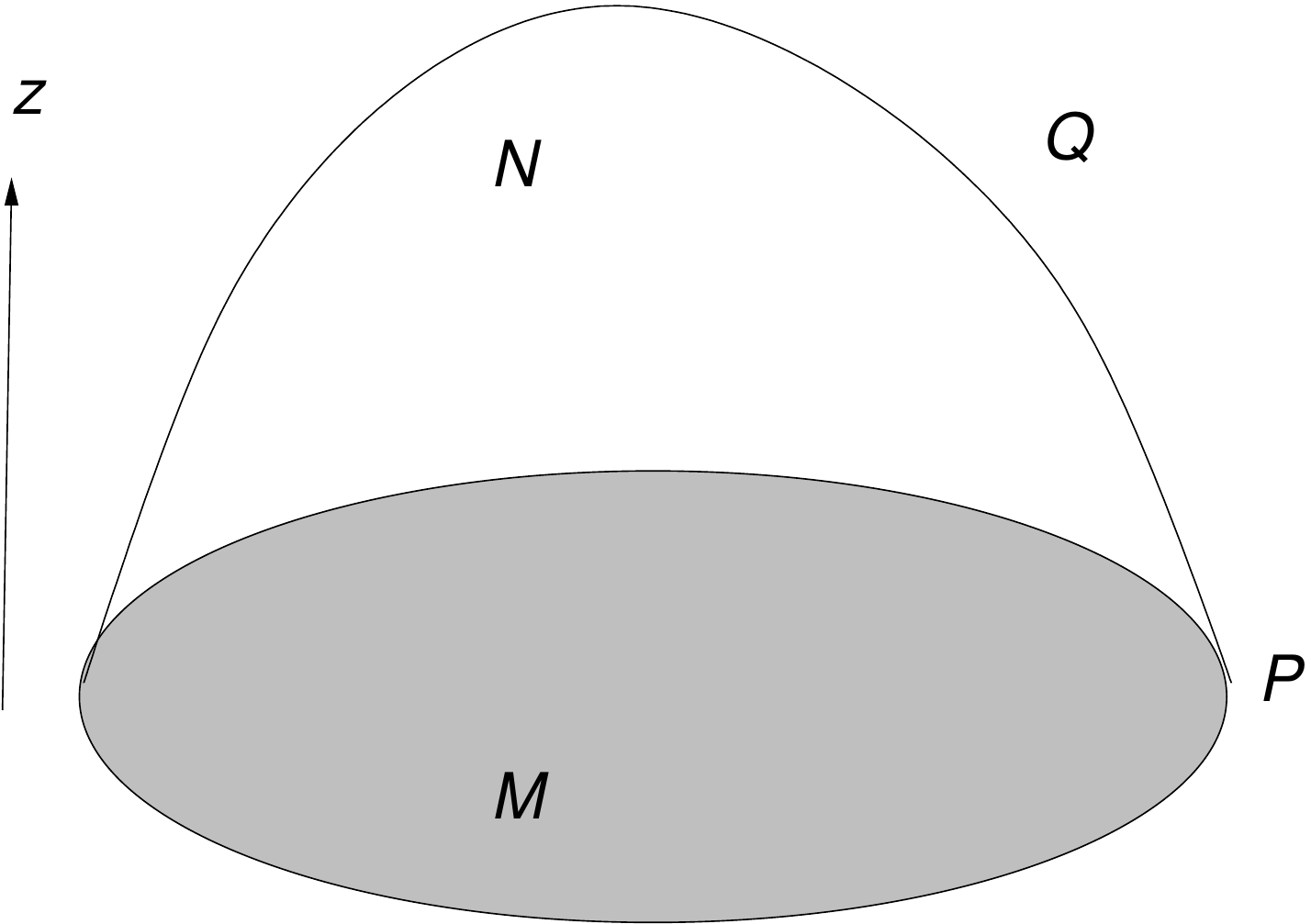}
\caption{BCFT on $M$ and its dual $N$}
\end{figure}

To make progress in this front, we find that one can
instead consider an expansion in
powers of small
%cr3 extrinsic curvature $k_{ab}$
derivatives of the metric
and keep both the $z$ and $x$ dependence as exact to construct
a perturbative solution to the Einstein equation.
% newv6
%After
%c5 spending a
%long effort,
%cr3
For simplicity, we consider the case of $h_{ab} = \d_{ab}$ here.
The more general case of a nontrivial boundary metric can be
analysed. We comment on this in the supplementary information.  
We find useful to consider
the following metric ansatz
% newv6
%with a function $f$ which depend on $x,z$ in a particular manner:
\begin{eqnarray}\label{bulkmetric}
%cr1
  ds^2=\frac{dz^2+ dx^2+\left(\delta_{ab}-2x \bar{k}_{ab} f
  \right)dy^a dy^b}{z^2}+ \cdots, \;\;\;\;\;\;\;
\end{eqnarray}
%cr2 Here
with $f = f(x,z)$ a function such that $f(x,0)=1$. To find solution, let
us first consider the region $x\geq 0$ and consider the ansatz $f=
f(z/x)$.
% newv6
This ansatz plays an important role to solve (\ref{NBC}) for non-symmetric boundary with $\bar{k}_{ab}\ne 0$.
 For simplicity we consider a
traceless $k_{ab} =\bar{k}_{ab}$ extrinsic curvature here.
The solution for the general case is given in
the
%c5 appendix.
supplementary information.
Substituting (\ref{bulkmetric}) into Einstein equation
%cr1
and writing  $s:=z/x >0$,
we obtain at the order $O(k)$ a single equation
\begin{eqnarray}\label{EOM}
s \left(s^2+1\right) f''(s)-(d-1) f'(s)=0.
\end{eqnarray}
%cr1 
It has the solution 
%$f(s)=1-\a_1\frac{s^d \,
% {}_2F_1\left(\frac{d-1}{2},\frac{d}{2};\frac{d+2}{2};-s^2\right)}{d}.$
% newv1 put the formula into maintext
\begin{eqnarray}\label{solution}
f(s)=1-\a_1\frac{s^d \,
 {}_2F_1\left(\frac{d-1}{2},\frac{d}{2};\frac{d+2}{2};-s^2\right)}{d}.
\end{eqnarray}
%cr1
To obtain a solution of the Einstein equation for $x<0$, one may analytic continuate
\eq{solution} to the region $s<0$. However this solution while
continuous at $s=0$, is discontinuous at $x=0$ as the region near
$x=0$ is mapped to widely separated regions $s =\pm \infty$.
Another possibility is to first rewrite the expression \eq{solution} in terms of
$x$ and $z$, and then  analytic continuate the resulting function
$f(x,z)$ to the region $x<0$. In this way, we obtain a solution of the
Einstein equation that is continuous at $x=0$. For example, for $d=3$, we have
\begin{subequations} \label{solutions3d}
 \begin{align} 
& f(x,z) =1-\a_1 (\frac{z}{x}-g(x,z)), \\
& g(x,z) =  \frac{\pi }{2}-2
\tan^{-1}\left(x/(z+\sqrt{z^2+x^2})\right).
\end{align}
  \end{subequations}
%cr1 
Let us make some comments. 1. For general $d$,
%cr1 it is remarkable that although $f(x,z)$ diverges at $x=0$,
the
 perturbation $2x \bar{k}_{ab} f(x,z)$ is finite which
 shows that
 %c4 it
 \eq{EOM} is a well-defined
 %c4 solution.
 metric. 2. Note that formally one can
 expand $f$ as a power series of $z$ and interpret that as a FG
 expansion
 %c5 for
 of the metric \eq{bulkmetric}. However the series does not
 converge whenever $x< z$. Therefore for the boundary ($x \to 0$) physics we are
 interested in, it is necessary to use the exact solution without
 performing the FG expansion.
 3. The perturbative background \eq{bulkmetric}, \eq{solution} to the Einstein
 equation is
 %c5 one of the main results of this letter 
 an interesting result which
 may be useful for other studies as well.

So far the coefficient $\a_1$ is arbitrary. If we now consider \eq{NBC}
in this background, we find that one
can solve the embedding function of $Q$ as $x=-\sinh(\rho) z+ O(k^2)$
provided that $\a_1$ is fixed at the same time.
% newv6
Please see the supplementary information for more details.
See Table \ref{table3} for
%c5
values of $\a_1$ obtained from holography,
%newv2
where we have re-parametrized $T=(d-1)\tanh \rho$ and
$\theta=\frac{\pi}{2}+2 \tan ^{-1}\left(\tanh \left(\frac{\rho
}{2}\right)\right)$ is the angle between
%cr1
$M$ and the bulk boundary $Q$.
%cr1 and AdS boundary $M$.
Using
\eq{bulkmetric}, \eq{solution}, we can derive the holographic stress
tensor \cite{deHaro:2000vlm}
\begin{eqnarray}\label{stresstensorsecond}
\ T_{ij}=\lim_{z\to 0} d \frac{\delta g_{ij}}{z^d}=2 \a_1 \frac{
  \bar{k}_{ij} }{x^{d-1}} +O(k^2),
\end{eqnarray}
which takes the expected form
\eq{SolutionTij1}. According to
\cite{deHaro:2000vlm}, $T_{ij}$ (\ref{stresstensorsecond})
automatically satisfy the traceless and divergenceless conditions
(\ref{stresstraceless}). 
% newv2
Note that in general the stress tensor (\ref{stresstensorsecond}) also
contains contributions
from $g_{ij}|_{z=0}$ in even dimensions
\cite{deHaro:2000vlm}. However, these contributions are finite, so we
can ignore them without loss of generality since we focus on only the
divergent parts in this letter.

Similarly, we can work out the next order solutions to both
%c4
the Einstein
equation and \eq{NBC}, and then derive the stress tensor
%c4 of
up to the order
$O(k^2)$ by applying the formula \eq{stresstensorsecond}.  See the
appendix for details. It turns out that the holographic stress tensor
takes exactly the expected expression
%c1 \eq{SolutionTij1}, \eq{SolutionTij2}, \eq{SolutionTij3}
\eq{solnT} with the
coefficients listed in Table \ref{table3}.
\begin{table}[ht]
\caption{Casimir coefficients for holographic stress tensor}
\begin{center}
    \begin{tabular}{| c | c | c | c |  c |  c |}
    \hline
     & $\a_1$ & $\b_1$ & $\b_2$  & $\b_3$ & $\b_4$\\ \hline
    3d   & $\frac{-1}{\theta}$ & 0 & 0 &
    $\frac{-2}{\theta}$ &0 \\ \hline 
    4d   & $\frac{-1}{2(1+\tanh \rho )}$ &
    $\frac{ \tanh \rho  }{ \tanh \rho 
      +1}$ & 0 &
    $\frac{5+4  \tanh \rho  }{-6 (1+ \tanh \rho)}$&
    $\frac{ \tanh \rho }{\tanh \rho 
      +1}$\\
    \hline
    \end{tabular}
\end{center}
\label{table3}
\end{table}
These coefficients indeed satisfy the relations
%c1(\ref{3dparameters},\ref{4dparameters1}) provided that
\eq{3dparameters}, \eq{4dparameters1}
%c3
provided the boundary central charges are given by \cite{note}
\begin{subequations}
  \label{ndcharges}
\begin{align}
&b_2=\frac{1}{\theta},  \label{3dhocharge}\\
&b_3= \frac{1}{1+\tanh \rho}-\frac{1}{3},\quad
  b_4=\frac{-1}{1+\tanh \rho},    \label{4dhochargebird1}
\end{align}
\end{subequations}
for 3d and 4d respectively.
%c3
Since we have
%c4 much more relations ($8\sharp$) than unknown variable ($3\sharp$),
many more relations (8) than unknown variables (3),
this is a non-trivial check of the  universal relations
\eq{3dparameters}, \eq{4dparameters1} as well
as for the holographic proposal \eq{NBC}.
In fact, the central charges (\ref{3dhocharge},\ref{4dhochargebird1}) can
be independently derived from the logarithmic divergent term of action
by using the perturbation solution of order $O(k^{d-1})$. One can
% c3 also
consider general boundary conditions by adding intrinsic curvatures
on $Q$ \cite{Chu:2017aab}. In this case the boundary central charges
change but the relations
\eq{3dparameters}, \eq{4dparameters1}
remain
the same. We can also reproduce these relations in the holographic
model \cite{Miao:2017gyt,Chu:2017aab}.
%c1 when taking into account the contributions on $Q$ carefully
%\cite{newwork}.
These are all strong supports for the universal relations
\eq{3dparameters}, \eq{4dparameters1}.
%cr1 and the holographic BCFT proposals of \cite{Takayanagi:2011zk} and
%cr1 \cite{Miao:2017gyt,Chu:2017aab}.
%cr1
The fact that the both  the holographic models
of \cite{Takayanagi:2011zk} and ours \cite{Miao:2017gyt,Chu:2017aab}
verify the universal relations \eq{3dparameters}, \eq{4dparameters1} 
suggests that both proposals are consistent holographic models of BCFT. 
We remark that in general there could be more than one self-consistent boundary
conditions for a theory \cite{Song:2016pwx} and so there is 
no contradiction between  \cite{Takayanagi:2011zk} and
\cite{Miao:2017gyt,Chu:2017aab}. This is supported by the fact that
the two holographic models gives different  boundary
central charges despite the same universal relations are satisfied.
%cr1 In particular the central charges (\eq{3dhocharge}, \eq{4dhocharge}) of 
% the holographic model \cite{Takayanagi:2011zk} and these of
% \cite{Miao:2017gyt,Chu:2017aab} agree
% in the limit $\rho \to -\infty$ for 3d
% BCFT and $\rho \to 0, -\infty$ for 4d BCFT, while disagree generally.

%c1 It is remarkable
%newv2
%We note that the boundary central charges
%\eq{3dhocharge}, \eq{4dhocharge} is discontinuous
%at $\rho=0$. This implies a first (resp. second) order
%phase transition for strongly coupled 3d (resp. 4d) BCFT
%c4 if
%as we change the boundary condition $\rho$
%c4 continuously
%in the holographic model \cite{Takayanagi:2011zk}. There is no such kinds of
%discontinuity in the holographic model of
%\cite{Miao:2017gyt,Chu:2017aab}.
%c1 should we mention this? the central charges in T are different for
%different methods

From holographic BCFT
\cite{Takayanagi:2011zk,Miao:2017gyt,Chu:2017aab}, we can
also gain some insight into the total energy. Applying the holographic
renormalization of BCFT \cite{Miao:2017gyt,Chu:2017aab}, we obtain the
total stress tensor:
\be\label{stressBCFTtotal}
T_{ij}=2 \a_1 \frac{\bar{k}_{ij}}{x^{d-1}}-\delta(x;P)  \frac{2\a_1}{d-2}  
\frac{\bar{k}_{ij}}{\epsilon^{d-2}}+O(k^2), \quad x \sim \epsilon.
\ee
Note that the first term, a local energy density, give rises to
a divergence in the total energy that cannot be canceled with any local
counterterm in the BCFT, but only with the inclusion of the second
term, a surface
counterterm as first constructed in \cite{Kennedy:1979ar}.
The surface counterterm is localized at the boundary surface $P$,
which has been shifted from $x=0$ to a position $x=\e$.
The requirement of finite energy 
fixes \cite{Kennedy:1979ar} the relative coefficients of the two terms in
\eq{stressBCFTtotal}.
%c13
Remarkably the holographic constructions
\cite{Takayanagi:2011zk,Miao:2017gyt,Chu:2017aab}
reproduce precisely
also the surface counter term with the needed coefficient to make the
total energy finite
%cr3 at least at the leading order
: $\int^{\infty}_{\epsilon} dx
T_{ij}=O(k^2)<\infty$, which agrees with the
results of \cite{Kennedy:1979ar,Dowker:1978md}.

%%%%%%%%%%%%%%%%%%%%%%%%%%%%%%%%%%%
\section{Conclusions and Discussions}
%%%%%%%%%%%%%%%%%%%%%%%%%%%%%%%%%%%

%c12
In this letter, we have shown that with the help of
an effective action description, the divergent parts of
the stress tensor of a BCFT is completely determined by the central charges of
the theory. The found relations between the Casimir
coefficients and the central charges are verified by free BCFT as well
as holographic models of BCFT. We propose that these relations
hold universally for any BCFT. Using the holographic models, we
also reproduce remarkably the precise surface counterterm that is needed
to render the total energy of the BCFT finite.

Our results are useful for the study of shape dependence of Casimir
effects 
% cc
\cite{Emig:2001dx,Schaden:2009zza,Rajabpour:2016iwf} and the theory of BCFT
\cite{Cardy:2004hm,McAvity:1993ue}. For Casimir
effects where there are spacetime on both sides of the boundary, it
has been
argued that the divergent stress tensor originates from the unphysical
nature of classical ``perfect conductor'' boundary conditions
\cite{Deutsch:1978sc}. In reality there would be an effective cut off
$\epsilon$ below which the short wavelength
vibrational modes do not ``see the boundary''.
%c1 From (\ref{SolutionTij2}), there could be a contribution to Casimir force
% \be\label{Casimirforce}
% F\sim \int_{P}\sqrt{h} \frac{2c_1}{d-2}\frac{\text{Tr}\bar{k}^2}{\epsilon^{d-2}}
% \ee
% Thus to produce large Casimir force, one should try to increase
% $\text{Tr}\bar{k}^2$ and decrease the effective cutoff.
%c4 While from the viewpoint of
However for BCFT where there is no spacetime outside the boundary,
the divergent one point function of stress tensor is expected and
physical. According to \cite{Cardy}, one can derive the one
point function of an operator in BCFT from the two point functions of
operators in CFT by using the mirror method. Since two point functions
are divergent when two points are approaching, it is not surprising
that the one point function of BCFT diverge near the boundary. This is
due to the interaction with the boundary, or equivalently, the mirror
image. Note that although the stress tensor diverges, the total energy
is finite. Thus BCFT is self-consistent.

Our discussions can be generalized to higher dimensions naturally. 
%newv6
Furthermore, our discussions also apply to defect conformal field
theory (DCFT) \cite{Billo:2016cpy} with general codimensions, which is
%cr2
a problem of great interest. For example, the case of
codimension 2 DCFT is related to the shape dependence of R\'enyi
entropy
%cr3 \cite{Lewkowycz:2014jia,Dong:2016wcf,Chu:2016tps,Bianchi:2016xvf}.
\cite{Lewkowycz:2014jia,Dong:2016wcf,Chu:2016tps,Bianchi:2016xvf,Bianchi:2015liz,Balakrishnan:2016ttg}.
%newv2
It is interesting to see whether the spirit of this letter can apply to general QFT. 
%It is remarkable that the main conclusion of this letter, that the
%divergent part of renormalized stress tensor can be completely fixed
%by Weyl
%anomaly, apply to not only BCFT but also more general QFT. See supplementary information for example.
It is also very interesting to generalize and apply the techniques of the
holographic
models to study the expectation value of current in boundary systems,
e.g. edge current of topological materials.

 \section*{Acknowledgements}
 We thank John Cardy,
 WuZhong Guo,
Hugh Osborn and Douglas Smith
for useful discussions and comments.
This work is supported in part by 
%the National Center of Theoretical Science
NCTS and the grant MOST
105-2811-M-007-021 of the Ministry of
Science and Technology of Taiwan.

\appendix

\section{{Solutions to holographic BCFT}}
Here we give details about solutions to the Einstein equations and the
boundary conditions \eq{NBC} to the next order
%cr3 $O(k^2)$
in derivative expansion of the boundary metric
%cr3
(i.e. $O(k^2)$ in the case of a flat boundary metric $h_{ab} = \d_{ab}$).
Consider the following ansatz
%cr1
for $x>0$,
\begin{eqnarray}\label{bulkmetrick2}
&& ds^2=\frac{1}{z^2}\Big{[} dz^2+ \big(1+x^2
  X(\frac{z}{x})\big)dx^2  \nonumber \\
&& +\big(\delta_{ab}-2x \bar{k}_{ab} f(\frac{z}{x})-2x \frac{k}{d-1}
  \delta_{ab}  + x^2 Q_{ab}(\frac{z}{x}) \big)dy^a dy^b\Big{]}  \nonumber\\
&&  +O(k^3),
\end{eqnarray}
where
%c4 $k$ denotes the trace of the extrinsic curvature $k_{ab}$ and
the functions $ X(\frac{z}{x})$ and $ Q_{ab}(\frac{z}{x})$ are of
order $O(k^2)$. We
%c4 set
require that
\be \label{fXQ}
f(0)=1,\quad X(0)=0, \quad Q_{ab}(0)=q_{ab}
\ee
so
that the metric of BCFT takes the form (\ref{BCFTmetric0})  in Gauss
normal coordinates.

\subsection{3d BCFT}
%c4 Now we
Let us
first study the case $d=3$. The generalization
to higher dimensions is straightforward. For simplicity, we further
set
%c6 \begin{eqnarray}\label{kqab}
$k_{ab}=\text{diag}(k_1, k_2),   q_{ab}=\text{diag}(q_1, q_2)$,
%c6 \end{eqnarray}
where $k_a, q_a$ are constants. Substituting (\ref{bulkmetrick2}) into the
Einstein equations,
%c6
and using  \eq{fXQ} to fix the integral constants,
%cr1 
we obtain \eq{solution} and 
  \bea \label{Qabapp1}
&&f(s) =1-\a_1 (s-g(s))\nonumber\\
&&Q_{11}(s)=\frac{1}{8}[ 4 q_1 \left(s^2+2\right)-\a_1^2
    \left(k_1-k_2\right){}^2 \left(s^2-3\right) g(s)^2\nonumber\\
&&-2 \a_1^2 \left(k_1-k_2\right){}^2 \log \left(s^2+1\right)+s \left(5
    \a_1^2 \left(k_1-k_2\right){}^2 s+4 \a_2\right)\nonumber\\
&&+s \left(2 \a_1 \left(-5 k_1^2+8 k_2 k_1+k_2^2\right)-4 s
    \left(k_1^2-k_2 k_1-k_2^2+q_2\right)\right)\nonumber\\
&&-2 g(s) \left(\a_1 k_1^2 \left(3 \a_1 s+s^2-5\right)+2 \a_2
    \left(s^2+1\right)\right)\nonumber\\
&&-2 \a_1 g(s) \left(k_2^2 \left(3 s \left(\a_1+s\right)+1\right)+2 k_1
    k_2 \left(4-3 \a_1 s\right)\right) ],\nonumber\\
&&Q_{22}(s)=\frac{1}{8}[ 4 q_2 \left(s^2+2\right)-\a_1^2
    \left(k_1-k_2\right){}^2 \left(s^2-3\right) g(s)^2\nonumber\\
&&+s \left(5 \a_1^2 \left(k_1-k_2\right){}^2 s-4 \a_2\right)-2 \a_1^2
    \left(k_1-k_2\right){}^2 \log \left(s^2+1\right)\nonumber\\
&&+s \left(4 s \left(k_1^2+k_2 k_1-k_2^2-q_1\right)-2 \a_1
    \left(k_1^2-4 k_2 k_1+7 k_2^2\right)\right)\nonumber\\
&&+2 g(s) \left(2 \a_2 \left(s^2+1\right)-\a_1 k_1^2 \left(3 \a_1
    s+s^2-1\right)\right)\nonumber\\
&&+2 \a_1 g(s) \left(k_2^2 \left(-3 \a_1 s+s^2+7\right)+2 k_1 k_2
    \left(3 \a_1 s+2 s^2-2\right)\right)],\nonumber\\
&&X(s)=\frac{1}{4}[-\a_1^2 \left(k_1-k_2\right){}^2 s^2 \log
    \left(s^2+1\right)-2 \a_1 \left(k_1-k_2\right){}^2 s\nonumber\\
&&+\a_1 \left(k_1-k_2\right){}^2 g(s) \left(\a_1 \left(s^2+1\right)
    g(s)+2 s \left(s-\a_1\right)+2\right)\nonumber\\
&&+s \left(\a_1^2 \left(k_1-k_2\right){}^2 s-2 s \left(k_1^2+k_2
    k_1+k_2^2-q_1-q_2\right)\right)],
\end{eqnarray}
%newv6
where $s=z/x$ and $g(s)=\frac{\pi }{2}-2\tan^{-1}\left(1/(s+\sqrt{s^2+1})\right)$.
%newv2
%where $g(s)=g(z/x)=\frac{\pi }{2}-2 \cot^{-1}\left(\sqrt{s^2+1}+s\right)$.
%cr1 where $g(s)=g(z/x)=\frac{\pi }{2}-2
%\tan^{-1}\left(x/(z+\sqrt{z^2+x^2})\right)$ which is smooth at $x=0$.
A continuous solution of the Einstein equations is obtained by first
rewriting \eq{bulkmetrick2} as function of $x$ and $z$ and then
analytic continutate to the $x<0$ region. 
% newv6
In this way, we get smooth $g(z,x)$ as (\ref{solutions3d}).
The solution is parametrized by two free parameters $\a_1$ and $\a_2$.

Next we solve \eq{NBC} for the embedding function of $Q$ in the above
background.
We obtain, for $d=3$, the results
\begin{eqnarray}\label{3dQapp1bird1}
x=-\sinh(\rho) z+\frac{k  \cosh ^2\rho  }{2 (d-1)} z^2 +c_3 z^3+O(k^3)
\end{eqnarray}
with $c_3$ given by
\begin{eqnarray}\label{c3Qapp1}
&&c_3=- \frac{\sinh \rho }{24}\Big[7 k_1^2+4 k_2 k_1+7 k_2^2-4
    \left(q_1+q_2\right) \nonumber\\
&&+\left(5 k_1^2+2 k_2 k_1+5 k_2^2-2 \left(q_1+q_2\right)\right) \cosh
    (2 \rho )\nonumber\\
    &&+\a_1^2 \left(k_1-k_2\right){}^2 \left((2 +\cosh (2 \rho ) )\log(
    \coth ^2 \rho )-1\right) \Big].\;\;\;
\end{eqnarray}
The boundary conditions (\ref{NBC}) also restrict solutions
(\ref{solutions3d}) and fix the integral constants to be
% newv2
%\begin{eqnarray}\label{c1c2Q}
%\a_1=\frac{1}{\tan ^{-1}(\text{csch}\rho )},\  \ \a_2=-\frac{\a_1}{2}k^2.
%\end{eqnarray}
\begin{eqnarray}\label{c1c2Qapp1}
\a_1=\frac{-1}{\theta},\  \ \a_2=-\frac{\a_1}{2}k^2,
\end{eqnarray}
where $\theta=\frac{\pi}{2}+2 \tan ^{-1}\left(\tanh \left(\frac{\rho
}{2}\right)\right)$ is the angle between
%cr1 
$M$ and the bulk boundary $Q$. It should be mentioned that, following
our method, the above $\alpha_1$ is independently obtained in a recent
paper \cite{Seminara:2017hhh}.
%cr1
The derivation of (\ref{3dQapp1bird1})-(\ref{c1c2Qapp1}) is straightforward.
For simplicity, let us first
focus on the leading order
$O(k)$ term. From  dimensional analysis, the embedding
function of $Q$ takes the form
$x=-\sinh(\rho) z+c_2 k z^2+O(k^2)$ with $c_2$ a dimensionless
constant. Substituting the metric \eq{bulkmetrick2} and the embedding
function of $Q$ into the conditions (\ref{NBC}), we get two
independent equations at order $O(k)$
\begin{eqnarray}\label{EOMc1metric}
&&\text{sech}^5(\rho ) (-8 c_2+\cosh (2 \rho)+1) k=0,\nonumber\\
&&\left(\alpha _1 \cosh ^2(\rho ) \left(4 \tan ^{-1}\left(\tanh
  \frac{\rho }{2}\right) +\pi\right)+8 c_2\right)
  \bar{k}_{ab}=0.\nonumber
\end{eqnarray}
Solving the above equations, we obtain $c_2$ and $\a_1$ as shown in
\eq{3dQapp1}, \eq{c1c2Qapp1}. Similarly, we obtain $c_3$ and $\a_2$
from (\ref{NBC}) at order $O(k^2)$.
It is remarkable that the
conditions (\ref{NBC}) fix the bulk metric and embedding function of
$Q$ at the same time.

% newv6
Substituting \eq{solutions3d}, \eq{bulkmetrick2},\eq{Qabapp1}, \eq{c1c2Qapp1} into
(\ref{stresstensorsecond}), we obtain the holographic stress
tensor 
\begin{eqnarray}\label{tij3dBCFT}
T_{ij}=\text{diag}\{&&\frac{\a_1  (k_1-k_2)^2}{x}, \frac{\alpha _1
  (k_1-k_2)}{x^2}-\frac{3 \alpha _1 (k_1-k_2){}^2}{2 x},\nonumber\\
 &&\frac{\alpha _1 (k_2-k_1)}{x^2}-\frac{3 \alpha _1 (k_1-k_2){}^2}{2 x}\}.
\end{eqnarray}
It is remarkable that all  the $q_a$ dependence got cancelled away 
and the stress tensor (\ref{tij3dBCFT}) takes exactly the expected form
%c1 \eq{SolutionTij1}, \eq{SolutionTij2},\eq{SolutionTij3})
\eq{solnT} with coefficients as listed in Table \ref{table3}.
% newv6
Recall that $k_{ij}$ in \eq{solnT} is actually a tensor defined at $x$
instead of the boundary $x=0$. It can be obtained from parallel
transport of the extrinsic curvature at $x=0$, i.e.,
$\bar{k}_{ij}(x)=g_i^{i'}g_j^{j'}\bar{k}_{i'j'}(0)=\bar{k}_{ij}(0)-2x
k_{(i}^l\bar{k}_{j)l}+O(x^2)$\cite{Deutsch:1978sc}.

Further generalization of the our above results is possible.
Let us discuss briefly the case of non-constant metric $h_{ij}(y)$ and
extrinsic curvature $k_{ij}(y)$. In this case, $T_{ij}$ will include
non-diagonal parts generally. These non-diagonal parts obey
(\ref{SolutionTij2}) trivially, since by definition
(\ref{stresstensorsecond}) $T_{ij}$
automatically satisfy the traceless and divergenceless conditions
(\ref{stresstraceless}), which fixs the non-diagonal parts of stress tensor as (\ref{SolutionTij2}) completely. 

%cr1
Another generalization is to have
more general boundary conditions of holographic BCFT by
adding intrinsic curvatures on $Q$ \cite{Chu:2017aab}. For example,
we consider
\begin{eqnarray}\label{actiongeneral}
I=\int_N \sqrt{G} (R-2 \Lambda) +2\int_Q \sqrt{\gamma} (K-T-\lambda R_Q),\;\;\;
\end{eqnarray}
with the Neumann boundary condition
%c1 insert lambda below
\begin{eqnarray}\label{NBCgeneral}
K_{\alpha\beta}-(K-T-\lambda R_Q)\gamma_{\alpha\beta}-2 \l R_{Q\alpha\beta}=0.
\end{eqnarray}
Substituting the solutions (\ref{solutions3d}) into
(\ref{NBCgeneral}), we can solve the embedding function of $Q$ as
(\ref{3dQapp1bird1}) but with different parameter $c_3$ and
different integration constants
\begin{eqnarray}\label{c1c2Qgeneral}
  \a_1&=&\frac{1}
    {2 \lambda  \rm{sech} \rho/\left(1-2 \lambda  \tanh \rho \right) -
      \theta},\nonumber\\
\a_2&=&-\frac{\a_1}{2}k^2.
\end{eqnarray}
Here $T=2\tanh \rho+2 \lambda  \text{sech}^2(\rho )$.
From (\ref{stresstensorsecond}), we can derive the holographic stress
tensor which takes exactly the expected form
%c1 (\ref{SolutionTij1},\ref{SolutionTij2},\ref{SolutionTij3}).
\eq{solnT}.
It is
remarkable that although the central charge $b_2=-\a_1$ changes, the
relations \eq{3dparameters} remain invariant for holographic BCFT
with general boundary conditions. The above discussions can be
generalized to higher dimensions easily. The 4d solutions can be used
to confirm the universal relations
\eq{4dparameters1}.

\subsection{4d BCFT}

Now Let us
consider the case $d=4$. For simplicity, we also
set
%c6 \begin{eqnarray}\label{kqab}
$k_{ab}=\text{diag}(k_1, k_2,k_3),   q_{ab}=\text{diag}(q_1, q_2,q_3)$,
%c6 \end{eqnarray}
where $k_a, q_a$ are constants. Substituting (\ref{bulkmetrick2}) into the
Einstein equations,
%c6
and using  \eq{fXQ} to fix the integral constants,
%cr1 
we obtain
\begin{eqnarray}\label{f}
f(s)=1+2 \alpha _1-\frac{\alpha _1 \left(s^2+2\right)}{\sqrt{s^2+1}},
\end{eqnarray}
  \bea \label{Qab}
&&X(s)=\frac{1}{6} s^2 \left(2 \left(q_1+q_2+q_3\right)-3 (k_1 k_2+k_1 k_3+k_3 k_2)\right)-\frac{1}{3} \left(k_1^2+k_2^2+k_3^2-k_1 k_2-k_1 k_3-k_2 k_3\right) g_1(s),\nonumber\\
&&Q_{11}(s)=\frac{k_1^2 g_2(s)+k_2^2 g_3(s)+k_1 k_2 g_4(s)}{18 \left(s^2+1\right)^{3/2}},\nonumber\\
&&+\frac{1}{3 \sqrt{s^2+1}} \left[q_1 \left(2 s^2+\sqrt{s^2+1}+2\right)+q_2 \left(-s^2+\sqrt{s^2+1}-1\right)+3 \alpha_2 \left(\left(\sqrt{s^2+1}-2\right) s^2+2 \left(\sqrt{s^2+1}-1\right)\right)\right],\nonumber\\
&&Q_{22}(s)=\frac{k_2^2 g_2(s)+k_1^2 g_3(s)+k_2 k_1 g_4(s)}{18 \left(s^2+1\right)^{3/2}}\nonumber\\
&&+\frac{1}{3 \sqrt{s^2+1}} \left[q_2 \left(2 s^2+\sqrt{s^2+1}+2\right)+q_1 \left(-s^2+\sqrt{s^2+1}-1\right)+3 \alpha_3 \left(\left(\sqrt{s^2+1}-2\right) s^2+2 \left(\sqrt{s^2+1}-1\right)\right)\right],\nonumber\\
&&Q_{33}(s)=\frac{(k_1^2+k_2^2) g_5(s)+k_1 k_2 g_6(s)}{18 \left(s^2+1\right)^{3/2}}\nonumber\\
&&\ \ \ \ -\frac{(q_1+q_2) \left(s^2-\sqrt{s^2+1}+1\right)+3 \left(\alpha _2+\alpha _3\right) \left(\left(\sqrt{s^2+1}-2\right) s^2+2 \left(\sqrt{s^2+1}-1\right)\right)}{3 \sqrt{s^2+1}},\nonumber\\
\end{eqnarray}
where $g_i(s)$ are defined by
\begin{eqnarray}\label{ggg}
&&g_1(s)=\alpha _1 \left(\alpha _1 \left(8 \sqrt{s^2+1}+s^2 \left(\log \left(s^2+1\right)-4\right)-8\right)-2 s^2+4 \sqrt{s^2+1}-4\right)+s^2\nonumber\\
&&g_2(s)=12 \left(s^2+1\right) (-s^2+\sqrt{s^2+1}-1)+36 \alpha _1 \left(s^2+1\right) (-s^2+2 \sqrt{s^2+1}-2)\nonumber\\
&&\ \ \ \ \ -\alpha _1^2 \left(-86 \left(\sqrt{s^2+1}-1\right)+s^2 \left(22 s^2-71 \sqrt{s^2+1}+108\right)+6 \left(s^2+1\right)^{3/2} \log \left(s^2+1\right)\right)\nonumber\\
&&g_3(s)=-6 \left(s^2+1\right) \left(-s^2+\sqrt{s^2+1}-1\right)+6 \alpha _1 \left(s^2+1\right) \left(-s^2+2 \sqrt{s^2+1}-2\right)\nonumber\\
&&\ \ \ \ +\alpha _1^2 \left(14 \left(\sqrt{s^2+1}-1\right)+s^2 \left(2 s^2+11 \sqrt{s^2+1}-12\right)-6 \left(s^2+1\right)^{3/2} \log \left(s^2+1\right)\right)\nonumber\\
&& g_4(s)=-12 \left(s^2+1\right) \left(-s^2+\sqrt{s^2+1}-1\right)-30 \alpha _1 \left(s^2+1\right) \left(-s^2+2 \sqrt{s^2+1}-2\right)\nonumber\\
&&\ \ \ \ +\alpha _1^2 \left(22 s^4-86 \left(\sqrt{s^2+1}-1\right)+s^2 \left(108-71 \sqrt{s^2+1}\right)+6 \left(s^2+1\right)^{3/2} \log \left(s^2+1\right)\right)\nonumber\\
&& g_5(s)=-6 \left(s^2+1\right) \left(-s^2+\sqrt{s^2+1}-1\right)-6 \alpha _1 \left(s^2+1\right) \left(\left(2 \sqrt{s^2+1}-3\right) s^2+2 \left(\sqrt{s^2+1}-1\right)\right)\nonumber\\
&&\ \ \ \ +\alpha _1^2 \left(44 \left(\sqrt{s^2+1}-1\right)+8 s^2 \left(7 \sqrt{s^2+1}-9\right)-6 \left(s^2+1\right)^{3/2} \log \left(s^2+1\right)+s^4 \left(15 \sqrt{s^2+1}-28\right)\right)\nonumber\\
&& g_6(s)=3 \left(s^2+1\right) \left(\left(3 \sqrt{s^2+1}-8\right) s^2+8 \left(\sqrt{s^2+1}-1\right)\right)\nonumber\\
&&\ \ \ \ +12 \alpha _1 \left(s^2+1\right) \left(\left(\sqrt{s^2+1}-3\right) s^2+4 \left(\sqrt{s^2+1}-1\right)\right)\nonumber\\
&&\ \ \ \ +\alpha _1^2 \left(4 s^2 \left(\sqrt{s^2+1}-6\right)+28 \left(\sqrt{s^2+1}-1\right)+6 \left(s^2+1\right)^{3/2} \log \left(s^2+1\right)+s^4 \left(4-15 \sqrt{s^2+1}\right)\right).\nonumber\\
\end{eqnarray}
Note that since the full expressions of $Q_{ab}$ are too complicated, we only list the results with $k_3=q_3=0$ for $Q_{ab}$ in (\ref{Qab}). We want to stress that we focus on the general case with nonzero $k_3$ and $q_3$, we just do not list the full expressions for simplicity. 

The above solutions work well for $x>0$. 
A continuous solution of the Einstein equations is obtained by first
rewriting \eq{bulkmetrick2},\eq{f},\eq{Qab},\eq{ggg} as functions of $x$ and $z$ and then
analytic continutate to the $x<0$ region. In fact, we only need to replace all $\sqrt{1+s^2}=\sqrt{1+\frac{z^2}{x^2}}$ in \eq{f},\eq{Qab},\eq{ggg} by $\sqrt{x^2+z^2}/x$. One can check that after the analytic continutation, the metric \eq{bulkmetrick2} are solutions to Einstein equations for $x\in (-\infty,\infty)$. What is more, now it becomes continuous at $x=0$ (see $x f(s)$ as an example). 
The above solution is parametrized by three free parameters $\a_1$, $\a_2$ and $\a_3$.

Next we solve \eq{NBC} for the embedding function of $Q$ in the above
background.
We obtain, for $d=4$, the results
\begin{eqnarray}\label{3dQ}
x=-\sinh(\rho) z+\frac{k  \cosh ^2\rho  }{2 (d-1)} z^2 +c_3 z^3+O(k^3)
\end{eqnarray}
with $c_3$ given by
\begin{eqnarray}\label{c3Q}
c_3=\frac{-1}{288}  e^{-2 \rho } \sinh (\rho ) \left[ t_1 (k_1^2+k_2^2+k_3^2)+t_2 (k_1 k_2+k_1 k_3+k_2 k_3)+t_3 (q_1+q_2+q_3)\right],
\end{eqnarray}
with $t_i$ given by
\begin{eqnarray}\label{ti}
&&t_1=8+48 \sinh (2 \rho )+20 \sinh (4 \rho )+5 \log \left(\coth ^2(\rho )\right) \nonumber\\
&& \ \ \ \ \ \     +\cosh (4 \rho ) \left(\log \left(\coth ^2(\rho )\right)+20\right)+\cosh (2 \rho ) \left(6 \log \left(\coth ^2(\rho )\right)+44\right)\nonumber\\
&&t_2=16+24 \sinh (2 \rho )+4 \sinh (4 \rho )-5 \log \left(\coth ^2(\rho )\right) \nonumber\\
&& \ \ \ \ \ \     -\cosh (4 \rho ) \left(\log \left(\coth ^2(\rho )\right)-4\right)+\cosh (2 \rho ) \left(28-6 \log \left(\coth ^2(\rho )\right)\right)\nonumber\\
&&t_3=-16 e^{2 \rho } (\cosh (2 \rho )+2).\nonumber\\
\end{eqnarray}
The boundary conditions (\ref{NBC}) also fix all the integral constants of the solutions
\eq{bulkmetrick2},\eq{f},\eq{Qab},\eq{ggg}
\begin{eqnarray}\label{c1c2Q}
&&\a_1=\frac{-1}{2 (\tanh (\rho )+1)},\nonumber\\
&&\a_2=\frac{-1}{144 (\sinh (\rho )+\cosh (\rho ))^2}\big{[}35 k_1^2+25 k_2 k_1+25 k_3 k_1-37 k_2^2-37 k_3^2-11 k_2 k_3-24 q_1+12 q_2+12 q_3\nonumber\\
&&\ \ \ \ \ \ \ \ +4 \left(4 k_1^2-7 \left(k_2+k_3\right) k_1-5 k_2^2-5 k_3^2+2 k_2 k_3-6 q_1+3 q_2+3 q_3\right) \sinh (2 \rho )\nonumber\\
&&\ \ \ \ \ \ \ \ +3 \left(k_1^2-5 \left(k_2+k_3\right) k_1-7 k_2^2-7 k_3^2-k_2 k_3-8 q_1+4 q_2+4 q_3\right) \cosh (2 \rho )
 \big{]}\nonumber\\
&&\a_3 = \a_2 [k_1\leftrightarrow k_2, q_1\leftrightarrow q_2].
\end{eqnarray}
It should be mentioned that, following our method, the above $\alpha_1$ is independently obtained in a recent paper \cite{Seminara:2017hhh}, which exactly agrees with our results when using our notations. 
%cr1
The derivation of (\ref{3dQ})-(\ref{c1c2Q}) is straightforward.
For simplicity, let us first
focus on the leading order
$O(k)$ term. From  dimensional analysis, the embedding
function of $Q$ takes the form
$x=-\sinh(\rho) z+c_2 k z^2+O(k^2)$ with $c_2$ a dimensionless
constant. Substituting the metric \eq{bulkmetrick2} and the embedding
function of $Q$ into the conditions (\ref{NBC}), we get two
independent equations at order $O(k)$
\begin{eqnarray}\label{EOMc1metric}
&&\text{sech}^5(\rho ) (-12 c_2+\cosh (2 \rho )+1) k=0,\nonumber\\
&&4 k_1 \left(\alpha _1 (\sinh (2 \rho )+\cosh (2 \rho )+1)+6 c_2\right)\nonumber\\
&&-\left(k_2+k_3\right) \left(2 \alpha _1 (\sinh (2 \rho )+\cosh (2 \rho )+1)+3 (-8 c_2+\cosh (2 \rho )+1)\right)=0.
\end{eqnarray}
Solving the above equations, we obtain $c_2$ and $\a_1$ as shown in
\eq{3dQ}, \eq{c1c2Q}
 \begin{eqnarray}\label{c2a1}
c_2=\frac{\cosh ^2(\rho )}{6}, \ \  \a_1=\frac{-1}{2 (\tanh (\rho )+1)}.
\end{eqnarray}
Similarly, we can obtain $c_3$, $\a_2, \a_3$
from Neumann boundary
conditions (\ref{NBC}) at the next order $O(k^2,q)$.
It is remarkable that the
conditions (\ref{NBC}) fix the bulk metric and embedding function of
$Q$ at the same time.

% newv6
Substituting the solutions \eq{bulkmetrick2},\eq{f},\eq{Qab},\eq{ggg},\eq{c1c2Q} into
the formula
\begin{eqnarray}\label{stresstensorsecondapp}
\ T_{ij}=\lim_{z\to 0} d \frac{\delta g_{ij}}{z^d},
\end{eqnarray}
and noting BCFT is defined in $x\in [0,\infty)$, we obtain the holographic stress tensor with non-zero components given by
\begin{eqnarray}\label{tij4dBCFT}
&&T_{xx}=-\frac{k_1^2+k_2^2+k_3^2-k_1 k_2-k_1k_3+k_2 k_3}{3 x^2 (\tanh (\rho )+1)},\nonumber\\
&&T_{11}=-\frac{\left(2 k_1-k_2-k_3\right) }{3(1+\tanh \rho) x^3}\nonumber\\
&&+\frac{\cosh (\rho )-\sinh (\rho )}{18 x^2} \big{[} 2 k_1^2 (5 \sinh (\rho )+8 \cosh (\rho ))+\left(k_2^2+k_3^2\right) (7 \cosh (\rho )-5 \sinh (\rho ))\nonumber\\
&& \ \ \ \ \ \ +2 k_2 k_3 (\sinh (\rho )+4 \cosh (\rho ))-k_1 \left(k_2+k_3\right) (\sinh (\rho )+19 \cosh (\rho ))+3 \left(q_2+q_3\right) \sinh (\rho )-6 q_1 \sinh (\rho ) \big{]},\nonumber\\
&&T_{22}=T_{11}[k_1\leftrightarrow k_2, q_1\leftrightarrow q_2],\nonumber\\
&&T_{33}=T_{11}[k_1\leftrightarrow k_3, q_1\leftrightarrow q_3].
\end{eqnarray}
We can rewrite the above holographic stress tensor into convariant form:
\begin{eqnarray}\label{tij4dBCFT1}
T_{ij}=\frac{2\a_1 (\bar{k}_{ij}-2x k_{(i}^l\bar{k}_{j)l})}{x^3}+ \frac{\a_1 (n_i n_j-\frac{h_{ij}}{3}) \text{Tr}\bar{k}^2 }{x^2}+\frac{p_1 C_{ikjl}n^kn^l+ p_2 k \bar{k}_{ij}+p_3 (k_{il}k^{l}_j-\frac{1}{3}h_{ij}\text{Tr}k^2)}{x^2}
\end{eqnarray}
where $\bar{}$ means traceless parts,  $C_{ikjl}n^kn^l=-\frac{1}{2}\bar{q}_{ij}+\frac{1}{2}k \bar{k}_{ij}$, $\alpha_1$ is given by (\ref{c1c2Q}), $n_i=(-1,0,0,0)$, $h_{ij}=\text{diag}(0,1,1,1)$ and $p_i$ are given by
\begin{eqnarray}\label{pi}
p_1=p_3=\frac{\tanh (\rho )}{\tanh (\rho )+1}, \ p_2=\frac{-4 \tanh (\rho )-5}{6 (\tanh (\rho )+1)}.
\end{eqnarray}

Now let us turn to the field theoretical result of BCFT stress tensor (\ref{SolutionTij1},\ref{SolutionTij2},\ref{SolutionTij3}), which takes the form 
\begin{eqnarray}\label{tij4dBCFT2}
T_{ij}=\frac{2\a_1 (\bar{k}_{ij}-2x k_{(i}^l\bar{k}_{j)l})}{x^3}+ \frac{\a_1 (n_i n_j-\frac{h_{ij}}{3}) \text{Tr}\bar{k}^2 }{x^2}+\frac{\b_1 C_{ikjl}n^kn^l+ \b_3 k \bar{k}_{ij}+\b_4 (k_{il}k^{l}_j-\frac{1}{3}h_{ij}\text{Tr}k^2)}{x^2}\nonumber\\
\end{eqnarray}
Recall that $k_{ij}$ in  eqs.(\ref{SolutionTij1},\ref{SolutionTij2},\ref{SolutionTij3}) is actually a tensor defined at $x$
instead of the boundary $x=0$. It can be obtained from parallel
transport of the extrinsic curvature at $x=0$, i.e.,
$\bar{k}_{ij}(x)=g_i^{i'}g_j^{j'}\bar{k}_{i'j'}(0)=\bar{k}_{ij}(0)-2x
k_{(i}^l\bar{k}_{j)l}+O(x^2)$\cite{Deutsch:1978sc}. 

Comparing the holographic stress tensor (\ref{tij4dBCFT1}) with the field theoretical result (\ref{tij4dBCFT2}), we get
\begin{eqnarray}\label{bi}
\b_1=p_1=\frac{\tanh (\rho )}{\tanh (\rho )+1}, \b_3=\ p_2=\frac{-4 \tanh (\rho )-5}{6 (\tanh (\rho )+1)},\ \b_4=\ p_3=\frac{-4 \tanh (\rho )-5}{6 (\tanh (\rho )+1)}.
\end{eqnarray}
Now it is easy to check that the Casimir coefficients  $\a_1, \b_i$ indeed satisfy the universal relations 
\be\label{4dparameters1s}
\begin{array}{lll}
  \a_1=\frac{b_4}{2}, & \b_1=\frac{c}{2\pi^2}+b_4, &\b_2=0, \\
&   \b_3=2 b_3+\frac{13}{6}b_4, &  \b_4=-3 b_3-2 b_4. 
\end{array}
\ee
provided the boundary central charges are given by 
\begin{eqnarray}
b_3= \frac{1}{1+\tanh \rho}-\frac{1}{3},\quad
  b_4=\frac{-1}{1+\tanh \rho},    \label{4dhocharge}
\end{eqnarray}
%c3
Since we have
four relations and two unknown variables,
this is a non-trivial check of the  universal relations
 \eq{4dparameters1s}. Recall that in (\ref{4dparameters1s}) we have $c=2\pi^2$ for Einstein gravity ($16\pi G_N=1$). In fact, the central charges (\ref{4dhocharge}) can
be independently derived from the logarithmic divergent term of action
by using the perturbation solution of order $O(k^{3})$.  It should be mentioned that the holographic results do not test the relations $\b_2=0$, since in our setup we have $h_{ab}=\delta_{ab}$ and thus $\mathcal{R}_{ij}=0$ in the stress tensor eq.(4) of the revised letter. However, this is a trivial relation and there is no need to test it. From the conformal symmetry, \cite{Deutsch:1978sc} finds that there is no $\bar{R}_{ij}=\bar{\mathcal{R}}_{ij}+O(k^2,q)$ terms in the stress tensor. As a result, we must have $\b_2=0$. 

Further generalization of the our above results is possible.
Let us discuss briefly the case of non-constant metric $h_{ij}(y)$ and
extrinsic curvature $k_{ij}(y)$. In this case, $T_{ij}$ will include
non-diagonal parts generally. These non-diagonal parts obey
eq.(3b) of the revised letter trivially, since by definition
(\ref{stresstensorsecondapp}) $T_{ij}$
automatically satisfy the traceless and divergenceless conditions, which fixs the non-diagonal parts of stress tensor as eq.(3b) in the revised letter  completely. 

Now we have shown that the holographic BCFT indeed obeys the universal relations (\ref{4dparameters1},\ref{4dparameters1s})  between Casimir coefficients and central charges.

%%%%%%%%%%%%%%%%%%%%%%%%%%%%%%%%%%%%%%

%%%%%%%%%%%%%%%%%%%%%%%%%%%%%%%%%%%%%%

\end{document}